\documentstyle[12pt]{article}

\begin{document}

\begin{titlepage}
\vskip 2cm
\begin{center}
{\bf\Large A Generalized Dual Symmetry for Nonabelian Yang-Mills
Fields$^{*)}$}
\vskip 1cm
{\large Chan Hong-Mo}\\
\vskip .3cm
{\it Rutherford Appleton Laboratory,\\
Chilton, Didcot, Oxon, OX11 0QX, UK.}\\
\vskip .5cm
{\large J. Faridani}\\
\vskip .3cm
{\it Department of Physics, University of Toronto,\\
60 St. George St., Toronto, ON, M5S 1A7, Canada.}\\
\vskip .5cm
{\large Tsou Sheung Tsun}\\
\vskip .3cm
{\it Mathematical Institute, Oxford University,\\
24-29 St. Giles', Oxford, OX1 3LB, UK.}\\
\vskip .5cm
\end{center}
\begin{abstract}
It is shown that classical nonsupersymmetric Yang-Mills theory in 4 dimensions
is symmetric under a generalized dual transform which reduces to the usual
dual *-operation for electromagnetism.  The parallel phase transport
$\tilde{A}_\mu(x)$ constructed earlier for monopoles is seen to function also
as a potential in giving a full description of the gauge field, playing thus
an entirely dual symmetric role to the usual potential $A_\mu(x)$.  Sources
of $A$ are monopoles of $\tilde{A}$ and vice versa, and the Wu-Yang criterion
for monopoles is found to yield as equations of motion the standard Wong and
Yang-Mills equations for respectively the classical and Dirac point charge;
this applies whether the charge is electric or magnetic, the two cases being
related just by a dual transform.  The dual transformation itself is explicit,
though somewhat complicated, being given in terms of loop space variables of
the Polyakov type.
\end{abstract}
\vfill
$^{*)}$ Dedicated to the memory of Professor Sir Rudolf Peierls, 1907 - 1995.
\end{titlepage}

\section{Introduction}
It is well-known that pure electrodynamics is symmetric under the interchange
of
electricity and magnetism: $E \rightarrow -H, H \rightarrow E$, or equivalently
under the Hodge star operation:\footnote{In our convention, $g_{\mu\nu}
= (+,-,-,-), \ \epsilon_{0123} = 1$.}
\begin{equation}
\mbox{\mbox{}$^*\!$} F_{\mu\nu} = - \mbox{\small $\frac{1}{2}$}
   \epsilon_{\mu\nu\rho\sigma} F^{\rho\sigma}.
\label{Hodgestar}
\end{equation}
This has led to many interesting consequences which have always intrigued
physicists \cite{Wuyang}-\cite{Sen} and have recently again excited much
interest due to the work of Seiberg, Witten and many
others \cite{Seibwitten}-\cite{Aharony}.

In view of the importance of Yang-Mills theories to particle physics,
it is natural to ask whether a similar symmetry exists also for nonabelian
gauge fields.  This question can be asked at many different levels.
Recently, it is most often addressed at the level of quantum fields, where
the Yang-Mills theory is embedded in a larger theory, usually supersymmetric
and existing in a high-dimensional space-time, in which charges, whether
electric or magnetic, appear as 't Hooft-Polyakov solitons \cite{Seibwitten}.
Here, however, we adopt a minimalist approach and ask whether strictly
4-dimensional and nonsupersymmetric Yang-Mills theory may possess a dual
symmetry at the classical field and point charge level.  Since it is at
this level that the Maxwell theory exhibits the well-known dual symmetry,
it seems reasonable to ask first whether Yang-Mills theory might
possess a generalized version of this symmetry also at the same level.

If duality for nonabelian theories is taken to mean again just the Hodge
star operation (\ref{Hodgestar}), then the answer to the above question
is no.  The field tensor $F_{\mu\nu}$ in the pure Maxwell theory satisfies the
equations:
\begin{equation}
F_{\mu\nu}(x) = \partial_\nu A_\mu(x) - \partial_\mu A_\nu(x),
\label{fmunu}
\end{equation}
and:
\begin{equation}
\partial^\nu F_{\mu\nu}(x) = 0.
\label{gausslaw0}
\end{equation}
By virtue of (\ref{fmunu}), $F_{\mu\nu}$ then satisfies the Bianchi identity:
\begin{equation}
\partial^\nu \mbox{\mbox{}$^*\!$} F_{\mu\nu}(x) = 0.
\label{gausslawd}
\end{equation}
Moreover, because the Hodge star operation is reflexive:
\begin{equation}
\mbox{\mbox{}$^*\!$} (\mbox{\mbox{}$^*\!$} F_{\mu\nu}) = - F_{\mu\nu}
\label{starstar}
\end{equation}
the Maxwell equation of (\ref{gausslaw0}) can similarly be interpreted in
this abelian case as the Bianchi identity for $\mbox{\mbox{}$^*\!$}
F_{\mu\nu}$, which then implies by the Poincar\'e lemma that there exists
a potential $\tilde{A}_\mu$ such that:
\begin{equation}
\mbox{\mbox{}$^*\!$} F_{\mu\nu}(x) = \partial_\nu \tilde{A}_\mu(x)
   - \partial_\mu \tilde{A}_\nu(x).
\label{fdmunu}
\end{equation}
One sees therefore that $F_{\mu\nu}(x)$ and
$\mbox{\mbox{}$^*\!$} F_{\mu\nu}(x)$ satisfy formally the same equations,
or that electromagnetism is dual symmetric.  For the pure nonabelian theory
on the other hand, the Yang-Mills field tensor satisfies, in parallel to
(\ref{fmunu}) and (\ref{gausslaw0}) for the abelian case, the equations:
\begin{equation}
F_{\mu\nu}(x) = \partial_\nu A_\mu(x) - \partial_\mu A_\nu(x)
   + ig[A_\mu(x), A_\nu(x)],
\label{Fmunu}
\end{equation}
and:
\begin{equation}
D^\nu F_{\mu\nu}(x) = 0,
\label{Gausslaw0}
\end{equation}
where $D_\mu$ denotes the usual covariant derivative:
\begin{equation}
D_\mu = \partial_\mu - ig [A_\mu(x), \ \ \ ].
\label{covderiv}
\end{equation}
Although (\ref{Fmunu}) implies again the Bianchi identity:
\begin{equation}
D^\nu \mbox{\mbox{}$^*\!$} F_{\mu\nu}(x) = 0,
\label{Gausslawd}
\end{equation}
this is not the dual of (\ref{Gausslaw0}), since the covariant derivative
in (\ref{Gausslawd}) involves the potential $A_\mu(x)$ and not some ``dual
potential'' appropriate to $\mbox{\mbox{}$^*\!$} F_{\mu\nu}(x)$.  Furthermore,
the Yang-Mills equation (\ref{Gausslaw0}) itself can no longer be interpreted
as the Bianchi identity for $\mbox{\mbox{}$^*\!$} F_{\mu\nu}(x)$, nor does it
imply the existence of a ``dual potential'' $\tilde{A}_\mu(x)$ satisfying:
\begin{equation}
\mbox{\mbox{}$^*\!$} F_{\mu\nu}(x) \stackrel{?}{=} \partial_\nu
\tilde{A}_\mu(x)
 -\partial_\mu \tilde{A}_\nu(x) +
i\tilde{g}[\tilde{A}_\mu(x),\tilde{A}_\nu(x)],
\label{Fdmunu}
\end{equation}
in parallel to (\ref{Fmunu}).  Indeed, it has been shown by Gu and Yang
\cite{Guyang} that for certain cases of $F_{\mu\nu}(x)$ satisfying
(\ref{Gausslaw0}) there are no solutions for $\tilde{A}(x)$ in (\ref{Fdmunu}),
which result shows once and for all that dual symmetry of Yang-Mills theory
under the Hodge star operation does not hold.

However, it is not excluded that there may be a generalized dual transform
which reduces to the Hodge star in the abelian case but for which there is
still an electric-magnetic dual symmetry for nonabelian Yang-Mills theory.
In fact, in an earlier paper \cite{Chanftsou1}, we have already suggested a
generalized dual transform which was able to reproduce many of what
one may call the dual properties of the abelian theory though not as yet the
complete dual symmetry.  The missing link in the arguments there for obtaining
a nonabelian dual symmetry was again the existence or otherwise of a local dual
potential $\tilde{A}_\mu(x)$ for Yang-Mills fields.  Although a local quantity
$\tilde{A}_\mu(x)$ did appear which functioned as the parallel transport for
the phase of colour magnetic charges exactly as a dual potential should, we
were unable to show that this $\tilde{A}_\mu(x)$ can reproduce all field
quantities - meaning that it gives a complete description of the theory.
As a result of this failure our treatment there, though having some desirable
features, remained far from being dual symmetric.

What we shall do in this paper is to show that a generalized dual symmetry
does exist for nonabelian Yang-Mills theory, and that the dual phase transport
$\tilde{A}_\mu(x)$ introduced in \cite{Chanftsou1} does function also
as a dual potential in that it gives a full description of the theory and
plays an entirely dual symmetric role to the standard gauge potential
$A_\mu(x)$.  This result is achieved by writing down a dual transform
between two new sets of variables which allows us to reformulate the whole
theory in an explicitly dual fashion.  Indeed, although the new results
are derived on the basis of results obtained before, the new dual symmetric
formulation is so much neater than the old that we shall find it easier
to derive some of the old results again together with the new than to
refer back to the older derivations.  We shall therefore work throughout
with the new dual formulation and only return in the end to sort out the
relationship with the older treatment.

A dual symmetry for Yang-Mills fields means in particular that colour
electric charges (i.e. ordinary colour charges such as quarks) which are
usually taken to be sources of the Yang-Mills field can also be considered
as monopoles of the dual field in the same way as colour magnetic charges
are monopoles of the Yang-Mills field.  It follows therefore that electric and
magnetic charges, in nonabelian as in abelian theories, have basically
the same dynamics, namely that given by the standard Maxwell and Yang-Mills
equations, only formulated in a dual manner.  Furthermore, since the relation
here between the field and the dual field though somewhat complicated is
explicitly known, the result may have brought us one step nearer to
realizing the hope of obtaining the strong coupling limit of one formulation
from the weak coupling limit of its dual by making use of the generalized
Dirac condition:
\begin{equation}
g \tilde{g} = \frac{1}{2N}
\label{Diraccond}
\end{equation}
relating the magnitudes of electric and magnetic couplings for a theory with
gauge group $SU(N)$.

\setcounter{equation}{0}
\section{$E_\mu[\xi|s]$ as Variables}

In our previous paper \cite{Chanftsou1} on Yang-Mills duality we have relied
heavily on a loop space technique developed earlier, using the Polyakov
variables $F_\mu[\xi|s]$ to describe the gauge
field \cite{Polyakov,Chanstsou,Chantsou}.  These variables $F_\mu[\xi|s]$
take values in the gauge Lie algebra, depend on the parametrized loop
$\xi$ only up to the point on $\xi$ labelled by the value $s$ of this
parameter, and have only components transverse to the loop at that point.
They are known to give a complete description of the Yang-Mills theory
but are highly redundant as all loop variables are, and have to be
constrained by an infinite set of conditions which is most conveniently
stated as the vanishing of the loop space curvature: \cite{Chanstsou,Chantsou}
\begin{equation}
G_{\mu\nu}[\xi|s] = 0,
\label{GausslawL}
\end{equation}
where:
\begin{equation}
G_{\mu\nu}[\xi|s] = \delta_\nu(s) F_\mu[\xi|s] - \delta_\mu(s) F_\nu[\xi|s]
   + ig [F_\mu[\xi|s], F_\nu[\xi|s]],
\label{Gmunu}
\end{equation}
and $\delta_\mu(s)$ denotes the loop derivative $\delta/\delta \xi^\mu(s)$
at $s$.  One great virtue of $F_\mu[\xi|s]$ as variables is that they are gauge
independent apart from an innocuous $x$-independent gauge rotation at the
fixed reference point $P_0$ for the parametrized loops.

In discussing dual properties, however, it was found convenient to introduce
another set of quantities $E_\mu[\xi|s]$ which were defined as:
\begin{equation}
E_\mu[\xi|s] = \Phi_\xi(s,0) F_\mu[\xi|s] \Phi^{-1}_\xi(s,0),
\label{Emuxis}
\end{equation}
where:
\begin{equation}
\Phi_\xi(s_2,s_1) = P_s \exp ig \int_{s_1}^{s_2} ds A_\mu(\xi(s))
   \dot{\xi}^\mu(s)
\label{Phixis2s1}
\end{equation}
is the parallel phase transport from the point at $s_1$ to the point at
$s_2$ along the loop $\xi$.  Hence, in order to exhibit
more clearly the dual properties of the theory, it is our intention here
to adopt these $E_\mu[\xi|s]$ instead of $F_\mu[\xi|s]$ as field variables.
Our first task is to demonstrate that this is possible under conditions
which we shall have to specify.

Recall first that the Polyakov variable $F_\mu[\xi|s]$ is defined as:
\begin{equation}
F_\mu[\xi|s] = \frac{i}{g} \Phi^{-1}[\xi] \delta_\mu(s) \Phi[\xi],
\label{Fmuxis}
\end{equation}
where:
\begin{equation}
\Phi[\xi] = P_s \exp ig \int_0^{2\pi} ds A_\mu(\xi(s)) \dot{\xi}^\mu(s),
\label{Phixi}
\end{equation}
or $\Phi_\xi(2\pi,0)$ as defined in (\ref{Phixis2s1}), so that $F_\mu[\xi|s]$
can be pictured as in Figure \ref{Fmuxisfig}, where
\begin{figure}
\vspace{7cm}
\caption{Illustration for $F_\mu[\xi|s]$}
\label{Fmuxisfig}
\end{figure}
the $\delta$-function $\delta(s-s')$ inherent in our definition of the loop
derivative\footnote{For any functional $\Psi[\xi]$ of the parametrized loop
$\xi$, we defined \cite{Chanstsou} the loop derivative $\delta_\mu(s)
= \delta/\delta\xi^\mu(s)$ as:
\begin{equation}
\frac{\delta}{\delta\xi^\mu(s)} \Psi[\xi] = \lim_{\Delta \rightarrow 0}
   \frac{1}{\Delta} \{\Psi[\xi'] - \Psi[\xi]\},\nonumber
\end{equation}
with:
\begin{equation}
\xi'^\alpha(s') = \xi^\alpha(s') + \Delta \delta_\mu^\alpha \delta(s-s').
   \nonumber
\end{equation}
In case of ambiguity, $\Delta \delta(s-s')$ in the expression above for
$\xi'^\alpha(s')$ is replaced by a bump-function with width $\epsilon$ and
height $h$, and the limit $\epsilon \rightarrow 0$ with $\Delta = \epsilon h$
held fixed is taken first, to be followed by the limit $h \rightarrow 0$.}
$\delta_\mu(s)$ is represented in the figure as a bump function centred at
$s$ with width $\epsilon = s_+ - s_-$.  In the same spirit, the quantity
$E_\mu[\xi|s]$ defined in (\ref{Emuxis}) can be pictured as the bold curve
in Figure \ref{Emuxisfig}
\begin{figure}
\vspace{7cm}
\caption{Illustration for $E_\mu[\xi|s]$}
\label{Emuxisfig}
\end{figure}
where the phase factors $\Phi_\xi(s,0)$ in (\ref{Emuxis}) have cancelled parts
of the circuit in Figure \ref{Fmuxisfig}.  In contrast to $F_\mu[\xi|s]$,
therefore, $E_\mu[\xi|s]$ is dependent really only on a ``segment'' of the
loop $\xi$ from $s_-$ to $s_+$.

The reason for representing the $\delta$-function in Figures \ref{Fmuxisfig}
and \ref{Emuxisfig} as a bump function is that, as in most functional
formulations, our treatment here involves some operations with the
$\delta$-function which need to be ``regularized'' to be given a meaning.
Our procedure is to take first the $\delta$-function as a bump function
with finite width, and then afterwards take the appropriate zero width limit.
For example, we shall need later the loop derivative $\delta_\nu(s)$ of the
quantity $E_\mu[\xi|s]$ at the same value of $s$.  Clearly, a loop derivative
has a meaning only if there is a segment of the loop on which it can operate.
Therefore, to define this derivative, we shall first regard $E_\mu[\xi|s]$
as a segmental quantity dependent on the segment of the loop $\xi$ from
$s - \epsilon/2$ to $s + \epsilon/2$.  We then define the loop derivative
$\delta_\nu(s)$ using the normal procedure on this segment, and afterwards
take the limit $\epsilon \rightarrow 0$.  In case a repeated loop derivative
of $E_\mu[\xi|s]$ is required at the same $s$, then the $\delta$-function
inherent in the first derivative has again to be represented by a bump
function of finite width, say $\epsilon'$, so that the second derivative
can be defined on this segment of the loop.  Afterwards, we
take first the limit $\epsilon' \rightarrow 0$, and then the limit
$\epsilon \rightarrow 0$, in that order.  In view of these regularization
procedures, it is often convenient to picture the quantities
$F_\mu[\xi|s]$ and $E_\mu[\xi|s]$ as in Figures \ref{Fmuxisfig} and
\ref{Emuxisfig}.

To show now that $E_\mu[\xi|s]$ do constitute a valid set of variables for
a full description of the gauge field, we note first that by (\ref{Emuxis})
and (\ref{Fmuxis}) we have:
\begin{equation}
\delta_\nu(s') E_\mu[\xi|s] = \Phi_\xi(s,0) \{ \delta_\nu(s') F_\mu[\xi|s]
   + ig \theta(s-s') [F_\nu[\xi|s'], F_\mu[\xi|s]] \} \Phi_\xi^{-1}(s,0),
\label{difEmuxis}
\end{equation}
where $\theta(s)$ is the Heaviside $\theta$-function, so that:
\begin{equation}
G_{\mu\nu}[\xi|s] = \Phi_\xi^{-1}(s,0) \{\delta_\nu(s) E_\mu[\xi|s]
   - \delta_\mu(s) E_\nu[\xi|s]\} \Phi_\xi(s,0),
\label{GmunuinE}
\end{equation}
and the condition (\ref{GausslawL}) translated in terms of $E_\mu[\xi|s]$
reads as:
\begin{equation}
\delta_\nu(s) E_\mu[\xi|s] - \delta_\mu(s) E_\nu[\xi|s] = 0.
\label{GausslawE}
\end{equation}
Hence, since we already know that $F_\mu[\xi|s]$ constrained by
(\ref{GausslawL}) describes the gauge theory, we want now to show that
given a set of $F_\mu[\xi|s]$ satisfying (\ref{GausslawL}) we recover a set
of $E_\mu[\xi|s]$ satisfying (\ref{GausslawE}) and vice versa.

The direct statement is easy to see.  Given $F_\mu[\xi|s]$ satisfying
(\ref{GausslawL}), we know from the so-called Extended Poincar\'e lemma
derived in \cite{Chanstsou} that we can recover a local potential
$A_\mu(x)$, from which a parallel transport $\Phi_\xi(s,0)$
by (\ref{Phixis2s1}), and hence also an $E_\mu[\xi|s]$ by (\ref{Emuxis})
can be constructed.  This $E_\mu[\xi|s]$ will automatically satisfy
(\ref{GausslawE}) as we wanted.

What is less obvious is the converse statement, namely that given a set of
$E_\mu[\xi|s]$ satisfying (\ref{GausslawE}), one can also recover a set of
$F_\mu[\xi|s]$ satisfying (\ref{GausslawL}).  To see this, one notes first
that given (\ref{GausslawE}), it follows that there exists some $W[\xi|s]$
such that:
\begin{equation}
E_\mu[\xi|s] = \delta_\mu(s) W[\xi|s].
\label{Wxis}
\end{equation}
Indeed, if one writes symbolically:
\begin{equation}
W[\xi|s] = \int_{\xi_0(s)}^{\xi(s)} \delta \xi'^\mu(s) E_\mu[\xi'|s]
\label{defWxis}
\end{equation}
as a line integral with respect to $\delta{\xi}$ along some path from an
arbitrary point $\xi_0(s)$ to the given point $\xi(s)$ then a similar
argument as in the usual Stokes' theorem would imply by (\ref{GausslawE})
that $W[\xi|s]$ is in fact path-independent and depends only on the end-point
$\xi(s)$ as indicated.  Furthermore, the derivative of this integral $W[\xi|s]$
would give $E_\mu[\xi|s]$ as desired.  If we now take a product of these
$W$'s along the loop $\xi$, thus:
\begin{equation}
\Phi_\xi(s,0) = P_{s'} \prod_{s'=0 \rightarrow s} \{1 - ig W[\xi|s']\},
\label{Phixis0inW}
\end{equation}
it is seen to satisfy:
\begin{equation}
\Phi_\xi^{-1}(s,0) \delta_\mu(s') \Phi_\xi(s,0) = -ig \theta(s-s')
   \Phi_\xi^{-1}(s,0) E_\mu[\xi|s] \Phi_\xi(s,0).
\label{difPhixis0}
\end{equation}
Defining then:
\begin{equation}
F_\mu[\xi|s] = \Phi_\xi^{-1}(s,0) E_\mu[\xi|s] \Phi_\xi(s,0)
\label{FmuxisinW}
\end{equation}
with $\Phi_\xi(s,0)$ given in (\ref{Phixis0inW}), we have:
\begin{eqnarray}
\lefteqn{\delta_\nu(s) F_\mu[\xi|s] - \delta_\mu(s) F_\nu[\xi|s]=} \nonumber \\
  &&   \Phi_\xi^{-1}(s,0) \{\delta_\nu(s) E_\mu[\xi|s]
   - \delta_\mu(s) E_\nu[\xi|s]\} \Phi_\xi(s,0)
    -ig [F_\mu[\xi|s], F_\nu[\xi|s]],
\label{curlFmu}
\end{eqnarray}
i.e. (\ref{GmunuinE}), which by (\ref{GausslawE}) means that
$G_{\mu\nu}[\xi|s]$
vanishes, as required.

In the above argument, however, we have actually glossed over a rather
important point, namely that in writing (\ref{difPhixis0}) we have
used (\ref{Wxis}) in which, by our procedure detailed above, $W[\xi|s]$
ought first to be regarded as a ``segmental quantity'' depending on a
segment of $\xi$ with width $\epsilon = s_+ - s_-$, and only after the
loop differentiation has been performed is the segmental width $\epsilon$
to be taken to zero.  On the other hand, in defining $\Phi_\xi(s,0)$ in terms
of $W[\xi|s]$, one wants already in (\ref{Phixis0inW}) to take the limit
$\epsilon \rightarrow 0$.  To assert both statements therefore, we shall need
a composition law for $W$ which says that the factor $(1 - ig W[\xi|s])$ for
a small finite segment is in fact the same as the product of such factors
for those infinitesimal segments which make up this small finite segment.
That such a composition law holds can be seen by an argument parallel to
that given in \cite{Chanstsou} for deriving the composition
law for $\Phi[\xi]$ by writing it in terms of $F_\mu[\xi|s]$ as a surface
integral.  Here, the line integral in loop space (\ref{defWxis}) representing
$W[\xi|s]$ is also in fact a surface integral in ordinary space-time for
which a similar argument is seen to apply.

That being the case, we conclude that $E_\mu[\xi|s]$ constrained by
(\ref{GausslawE}) do constitute a valid set of variables for describing the
gauge field, which we shall adopt later for discussing its dual properties.
Note that, in contrast to the Polyakov variables $F_\mu[\xi|s]$, the variables
$E_\mu[\xi|s]$ are gauge dependent quantities and so, though more convenient
than $F_\mu[\xi|s]$ for studying duality, may not be so useful otherwise.
We note further that the fact we are able to recover from $E_\mu[\xi|s]$
satisfying (\ref{GausslawE}) the Polyakov variables $F_\mu[\xi|s]$
satisfying (\ref{GausslawL}) means also by the Extended Poincar\'e lemma
of \cite{Chanstsou} that there exists a local potential $A_\mu(x)$ such
that the parallel transport is indeed given by (\ref{Phixis2s1}). In turn,
this implies that:
\begin{equation}
\lim_{\epsilon \rightarrow 0} W[\xi|s] = \lim_{s_+ \rightarrow s_-}
   \frac{i}{g} \{\Phi_\xi(s_+,s_-) - 1\} = A_\mu(\xi(s)) \dot{\xi}^\mu(s),
\label{AmuxinW}
\end{equation}
and that:
\begin{equation}
\lim_{\epsilon \rightarrow 0} E_\mu[\xi|s] =F_{\mu\nu}(\xi(s))
\dot{\xi}^\nu(s),
\label{FmunuxinE}
\end{equation}
with $F_{\mu\nu}(x)$ given as usual in (\ref{Fmunu}) in terms of the
$A_\mu(x)$ defined in (\ref{AmuxinW}) above.  These two formulae will be
of use to us later.

\setcounter{equation}{0}
\section{Generalized Dual Transform}

As noted above in the Introduction, Hodge star duality does not lead to a
dual symmetry for nonabelian Yang-Mills theory.  We seek therefore a
generalized dual transform, if such exists, which may restore dual
symmetry to Yang-Mills theory.  The experience gained in earlier work leads
us to believe that such a transform is best written in terms of the
variables $E_\mu[\xi|s]$ introduced in the preceding section.

We seek a dual transform with the following 3 properties.  First, we want,
of course, that the new dual transform reduces back to the Hodge star
(\ref{Hodgestar}) for the abelian theory, but that it should not
do so for the nonabelian case or else the conclusion of Gu and Yang in
\cite{Guyang} would be violated.  Secondly, in order for the new transform
to qualify as a dual transform, we want it to be invertible in the sense
that, like the Hodge star, application of the transform twice should give
the identity, apart perhaps from a sign.  Thirdly, we want the transform
to be such that, again like the Hodge star in the abelian case, an electric
charge defined as a source of the direct field should appear as a monopole
of the dual field, while a magnetic charge defined as a source of the dual
field should appear as a monopole of the direct field.  This last property
seems to us to be the crucial feature which gives dual symmetry to the
abelian theory and which, we have reason to believe from past experience, may
give dual symmetry also to Yang-Mills fields.

Our suggestion is as follows.  Given a set of variables $E_\mu[\xi|s]$
describing the gauge field, we introduce a corresponding dual set of variables
$\tilde{E}_\mu[\eta|t]$ labelled by $\eta$ and $t$, where $\eta$ is just
another parametrized loop with parameter $t$ which are distinguished here
by different symbols from $\xi$ and $s$ for convenience.  For given
$\eta$ and $t$, $\tilde{E}_\mu[\eta|t]$ is defined as:
\begin{equation}
\omega^{-1}(\eta(t)) \tilde{E}_\mu[\eta|t] \omega(\eta(t)) = -\frac{2}{\bar{N}}
   \epsilon_{\mu\nu\rho\sigma} \dot{\eta}^\nu(t) \int \delta\xi ds
   E^\rho[\xi|s] \dot{\xi}^\sigma(s) \dot{\xi}^{-2}(s) \delta(\xi(s)-\eta(t)),
\label{newduality}
\end{equation}
where $\omega(x)$ is just a local rotational matrix allowing for the
freedom of transforming from the ``$U$''-frame in which direct quantities
like $E_\mu[\xi|s]$ are represented to a ``$\tilde{U}$''-frame in which
dual quantities like $\tilde{E}_\mu[\eta|t]$ are represented, and $\bar{N}$
a normalization factor (infinite) defined as: \cite{Chanstsou,Chantsou,
Chanftsou1}
\begin{equation}
\bar{N} = \int_0^{2\pi} ds \prod_{s' \neq s} d^4\xi(s').
\label{Nbar}
\end{equation}

As (\ref{newduality}) involves an implicit regularization procedure, i.e. a
fixed order in which various limits are taken, some explanation is in order.
The loop integral on the right-hand side of (\ref{newduality}), as for the
loop derivative discussed in the preceding section, needs a segment of the
loop $\xi$ on which to operate.  Hence, $E_\rho[\xi|s]$ has first again to be
regarded as a segmental quantity depending on a little segment of $\xi$ from
$s_-$ to $s_+$ whose width $\epsilon = s_+ - s_-$ is taken to zero only after
the integration has been performed.  In the same spirit, $\dot{\xi}(s)$ in
the integrand is meant to represent the quantity $(\xi(s_+)-\xi(s_-))/\epsilon$
which becomes the tangent to the loop $\xi$ at $s$ when $\epsilon \rightarrow
0$.  If one is interested only in the value of $\tilde{E}_\mu[\eta|t]$ and
not, say, in its derivatives, then $\tilde{E}_\mu[\eta|t]$ can be taken as
just a function of the point $\eta(t)$ labelled by $t$ on the loop $\eta$
and of the tangent $\dot{\eta}(t)$ to the loop at that point.  In that
case, the $\delta$-function $\delta(\xi(s)-\eta(t))$ on the right says
that the segment $\xi$ has to pass through at $s$ the point $\eta(t)$
but is otherwise freely integrated so that $\dot{\xi}(s)
= (\xi(s_+)-\xi(s_-))/\epsilon$ can have any direction relative to
$\dot{\eta}(t)$, except that the contribution to the integral vanishes when
$\dot{\xi}(s)$ is parallel to $\dot{\eta}(t)$ because of the
$\epsilon_{\mu\nu\rho\sigma}$ symbol in front.  However, if we wish to
evaluate the loop derivative $\delta_\alpha(t) = \delta/\delta \eta^\alpha(t)$
of $\tilde{E}_\mu[\eta|t]$ using the formula (\ref{newduality}), then
$\tilde{E}_\mu[\eta|t]$ itself has also to be regarded as a segmental
quantity depending on a segment of $\eta$ from $t_-$ to $t_+$ with width
$\epsilon' = t_+ - t_-$.  After the differentiation has been performed,
one can then take the limit $\epsilon' \rightarrow 0$, and our procedure
says that this limit should be taken before the limit $\epsilon \rightarrow 0$
for the integral.  That being the case, we may take $\epsilon' < \epsilon$,
and the $\delta$-function $\delta(\xi(s)-\eta(t))$ should now be interpreted
as saying that the segment $\xi$ coincides from $s = t_-$ to $s = t_+$ with the
segment $\eta$, but outside that interval is still freely integrated so that
$\dot{\xi}(s)$ can again have any direction relative to $\dot{\eta}(t)$.
Since the integral receives contributions only from $\xi$ segments with
$\dot{\xi}$ nonparallel to $\dot{\eta}$, we cannot take $\epsilon' = \epsilon$,
otherwise $\dot{\xi}(s) = \dot{\eta}(t)$ and the integral would vanish.

With these clarifications in the interpretation of the dual transform
(\ref{newduality}) let us now examine whether colour electric charges do
indeed appear as monopoles of the dual field $\tilde{E}_\mu[\eta|t]$, which
property, as stated above, we believe to be crucial for dual symmetry.  We
recall first that a colour electric charge is usually defined as a source of
the Yang-Mills field, namely a nonvanishing covariant divergence
$D^\nu F_{\mu\nu}(x)$.  Equivalently, according to Polyakov \cite{Polyakov},
it is a nonvanishing loop divergence $\delta^\mu(s) F_\mu[\xi|s]$ of the loop
variable $F_\mu[\xi|s]$.  Alternatively again, since (\ref{difEmuxis})
implies that:
\begin{equation}
\delta^\mu(s) E_\mu[\xi|s] = \Phi_\xi(s,0) \{\delta^\mu(s) F_\mu[\xi|s]\}
   \Phi_\xi^{-1}(s,0),
\label{divEinF}
\end{equation}
it also means a nonvanishing loop divergence $\delta^\mu(s) E_\mu[\xi|s]$
of the variable $E_\mu[\xi|s]$ adopted here.  On the other hand, a colour
magnetic charge defined as a monopole of the Yang-Mills field is characterized
most easily as a nonvanishing loop space curvature \cite{Chanstsou,Chantsou}
$G_{\mu\nu}[\xi|s]$ as defined in (\ref{Gmunu}), or alternatively, by
(\ref{GmunuinE}) in terms of $E_\mu[\xi|s]$, as a
nonvanishing ``curl'' $\delta_\nu(s) E_\mu[\xi|s] - \delta_\mu(s)
E_\nu[\xi|s]$.
By a monopole of the dual field $\tilde{E}$ we mean then a nonvanishing curl
$\delta_\nu(t) \tilde{E}_\mu[\eta|t] - \delta_\mu(t) \tilde{E}_\nu[\eta|t]$.
Hence to show that a colour electric charge is indeed a monopole of the dual
field, we need to show that a nonvanishing divergence of $E$ will lead to a
nonvanishing curl of the dual variable $\tilde{E}$ as defined by the dual
transform (\ref{newduality}).  The parallel for this in the abelian theory
is that an electric charge represented by the nonvanishing divergence
$\partial^\nu F_{\mu\nu}(x)$ of the Maxwell field can also be interpreted
as the violation of the Bianchi identity for the dual field
$\mbox{\mbox{}$^*\!$} F_{\mu\nu}(x)$, which signifies the presence of a
monopole in $\mbox{\mbox{}$^*\!$} F$.

That a nonvanishing divergence of $E$ would generally lead to a nonvanishing
curl of $\tilde{E}$ can be seen by direct computation.  From
(\ref{newduality}),
one can write:
\begin{eqnarray}
& & \epsilon^{\lambda\mu\alpha\beta} \delta_\lambda(t)
   \{\omega^{-1}(\eta(t)) \tilde{E}_\mu[\eta|t] \omega(\eta(t))\} \nonumber\\
    & = & -\frac{2}{\bar{N}} \epsilon^{\lambda\mu\alpha\beta}
   \epsilon_{\mu\nu\rho\sigma} \dot{\eta}^\nu(t) \int \delta\xi ds
   \{\delta_\lambda(s) E^\rho[\xi|s]\}
   \dot{\xi}^\sigma(s) \dot{\xi}^{-2}(s) \delta(\xi(s) - \eta(t)),
\label{curlEtilde1}
\end{eqnarray}
where, recalling from the above paragraph that in $\delta(\xi(s)-\eta(t))$
on the right, $\eta(t)$ is first to be interpreted as a little segment which
coincides with $\xi(s)$ for $s = t_- \rightarrow t_+$, we have put
$\delta_\lambda(t) = -\delta_\lambda(s)$ and then performed an
integration by parts with respect to $\delta\xi$.  Expressing next
$\epsilon^{\lambda\mu\alpha\beta} \epsilon_{\mu\nu\rho\sigma}$ as a combination
of Kronecker deltas and using the fact that segmental quantities, like loop
quantities, have only transverse loop derivatives so that both $\delta_\mu(s)
\dot{\xi}^\mu(s)$ and $\delta_\mu(t) \dot{\eta}^\mu(t)$ vanish, we
obtain for (\ref{curlEtilde1}):
\begin{eqnarray}
\hspace*{-5mm}& & \epsilon^{\lambda\mu\alpha\beta} \delta_\lambda(t)
   \{\omega^{-1}(\eta(t)) \tilde{E}_\mu[\eta|t] \omega(\eta(t))\} \nonumber \\
& = &\!\!\!\!-\frac{2}{\bar{N}} \int\delta\xi ds \{ \dot{\eta}^\beta(t)
   \dot{\xi}^\alpha(s) - \dot{\eta}^\alpha(t) \dot{\xi}^\beta(s) \}
   \delta_\rho(s) E^\rho[\xi|s] \dot{\xi}^{-2}(s) \delta(\xi(s)-\eta(t)).
\label{curlEtilde2}
\end{eqnarray}
On multiplying by $\mbox{\small $\frac{1}{2}$} \epsilon_{\mu\nu\alpha\beta}$,
we obtain:
\begin{eqnarray}
& & \omega^{-1}(\eta(t)) \{\delta_\nu(t) \tilde{E}_\mu[\eta|t]
   - \delta_\mu(t) \tilde{E}_\nu[\eta|t]\} \omega(\eta(t)) \nonumber \\
   & = & -\frac{1}{\bar{N}} \int \delta\xi ds \epsilon_{\mu\nu\alpha\beta}
   \{\dot{\eta}^\beta(t) \dot{\xi}^\alpha(s)- \dot{\eta}^\alpha(t)
   \dot{\xi}^\beta(s)\} \delta_\rho(s) E^\rho[\xi|s] \dot{\xi}^{-2}(s)
   \delta(\xi(s)-\eta(t)),\nonumber \\
&&
\label{curlEtilde}
\end{eqnarray}
where the factors $\omega^{-1}(\eta(t))$ and $\omega(\eta(t))$ can be taken
outside because loop derivatives vanish for local quantities.\footnote{Although
$\omega(\eta(t))$ does vary when $\eta$ is varied at $t$, its variation is
of measure zero compared with the variation of the loop so long as the
$\delta$-function in the definition of the loop derivative is given a
finite width, so that the derivative has to be assigned the value zero
for consistency with our standard procedure for resolving such ambiguities.}
One sees thus that the divergence of $E$ is indeed related to the curl of
$\tilde{E}$ and that an electric charge characterized by the nonvanishing
of the former will in general mean a monopole characterized by a nonvanishing
curl of the latter.  Conversely, if $\delta^\rho(s) E_\rho[\xi|s] = 0$ then
$\delta_\nu(t) \tilde{E}_\mu[\eta|t] - \delta_\mu(t) \tilde{E}_\nu[\eta|t]=0$,
or in other words the absence of sources in $E$ will
guarantee the absence of monopoles in $\tilde{E}$, which statement is in
fact what is needed for deriving dual symmetry, as we shall see later.

Next, we wish to check that (\ref{newduality}) reduces to the Hodge star
relation when the theory is abelian but not when the theory is nonabelian.
To see this, we let the segmental width of $\tilde{E}_\mu[\eta|t]$ in
(\ref{newduality}) go to zero so that we can use the formula (\ref{FmunuxinE})
to write the left-hand side in terms of local quantities:
\begin{equation}
\omega^{-1}(x) \tilde{F}_{\mu\nu}(x) \omega(x) = -\frac{2}{\bar{N}}
   \epsilon_{\mu\nu\rho\sigma} \int \delta\xi ds E^\rho[\xi|s]
   \dot{\xi}^\sigma(s) \dot{\xi}^{-2}(s) \delta(x - \xi(s)).
\label{reducedual1}
\end{equation}
We recall that our procedure is to do the integral before taking the width
of the segment in $E_\mu[\xi|s]$ to zero.  In other words, within the integral,
the loop $\xi$ can still vary by a $\delta$-functional bump as illustrated in
Figure \ref{reducedualfig} (a).  For such a $\xi$, $E_\mu[\xi|s]$, which is
\begin{figure}
\vspace{7cm}
\caption{Illustration for the Integrand in Dual Transform}
\label{reducedualfig}
\end{figure}
obtained by making a $\delta$-functional variation along the direction
$\mu$, will take on the shape depicted in Figure \ref{reducedualfig} (b).  This
last figure can be expressed as the product of three factors, namely Figures
\ref{reducedualfig}(c),(d),(e) in the order indicated.  In the abelian
theory, the ordering of the factors is unimportant so that the factors of
Figures (c) and (e) cancel in the limit when the segmental width
$\epsilon \rightarrow 0$, leaving only the factor of Figure (d), which can
as usual be expressed by (\ref{FmunuxinE}) as $F_{\mu\alpha}(\xi(s))
\dot{\xi}^\alpha(s)$, giving:
\begin{eqnarray}
\tilde{F}_{\mu\nu}(x) & = & -\frac{2}{\bar{N}} \epsilon_{\mu\nu\rho\sigma}
   \int \delta\xi ds F^{\rho\alpha}(\xi(s)) \dot{\xi}_\alpha(s)
   \dot{\xi}^\sigma(s) \dot{\xi}^{-2}(s) \delta(x-\xi(s)) \nonumber \\
   & = & - \mbox{\small $\frac{1}{2}$} \epsilon_{\mu\nu\rho\sigma}
   F^{\rho\sigma}(x),
\label{reducedual}
\end{eqnarray}
which is just the Hodge star relation if we identify $\tilde{F}_{\mu\nu}(x)$
with $\mbox{\mbox{}$^*\!$} F_{\mu\nu}(x)$.  On the other hand, for a nonabelian
theory, the factors of Figures \ref{reducedualfig} (c) and (e) cannot be
commuted through the factor of Figure (d) so that the above reduction to the
Hodge star relation will not go through.

Lastly, we wish to examine whether the dual transform (\ref{newduality})
is invertible.  From (\ref{newduality}) we can write:
\begin{eqnarray}
&  & \frac{2}{\bar{N}} \epsilon^{\alpha\beta\mu\lambda} \dot{\zeta}_\beta(u)
   \int \delta\eta dt \omega^{-1}(\eta(t)) \tilde{E}_\mu[\eta|t]
\omega(\eta(t))
   \dot{\eta}_\lambda(t)\dot{\eta}^{-2}(t) \delta(\eta(t)-\zeta(u)) \nonumber
\\
& = & -\frac{4}{\bar{N}^2} \epsilon^{\alpha\beta\mu\lambda}
   \epsilon_{\mu\nu\rho\sigma} \dot{\zeta}_\beta(u) \int \delta\eta dt
   \dot{\eta}_\lambda(t) \dot{\eta}^\nu(t) \dot{\eta}^{-2}(t)
   \delta(\eta(t)-\zeta(u)) \nonumber \\
&  & \int \delta\xi ds E^\rho[\xi|s] \dot{\xi}^\sigma(s)
   \dot{\xi}^{-2}(s) \delta(\xi(s)-\eta(t)).
\label{invertdual1}
\end{eqnarray}
By integrating first over all directions of $\dot{\eta}(t)$ which we recall
from the explanation given after (\ref{newduality}) is admissible, we
obtain a factor $\bar{N} \delta_\lambda^\nu/4$, so that the right-hand
side reduces to:
\begin{equation}
\frac{2}{\bar{N}} \{\delta_\rho^\alpha \delta_\sigma^\beta - \delta_\rho^\beta
   \delta_\sigma^\alpha\} \dot{\zeta}_\beta(u) \int \delta\xi ds E^\rho[\xi|s]
   \dot{\xi}^\sigma(s) \dot{\xi}^{-2} \delta(\xi(s)-\eta(t)).
\label{invertdual2}
\end{equation}
Using the argument in the paragraph above, one can show that the integral
in (\ref{invertdual2}) is antisymmetric in the indices $\rho$ and $\sigma$
giving then just twice the first term where, since $\dot{\zeta}$ and
$\dot{\xi}$ are no longer forbidden to be parallel, we may put them equal
using $\delta(\xi(s)-\zeta(u))$ so that the whole expression reduces to
just $E^\alpha[\zeta|u]$, giving:
\begin{equation}
\omega(\zeta(u)) E_\alpha[\zeta|u] \omega^{-1}(\zeta(u))
   = \frac{2}{\bar{N}} \epsilon_{\alpha\beta\mu\lambda}
   \dot{\zeta}^\beta(u) \int \delta\eta dt \tilde{E}^\mu[\eta|t]
   \dot{\eta}^\lambda(t) \dot{\eta}^{-2}(t) \delta(\eta(t)-\zeta(u)),
\label{invertdual}
\end{equation}
as required.

We have now shown that the generalized dual transform suggested in
(\ref{newduality}) does indeed have all the 3 properties that we desired.

\setcounter{equation}{0}
\section{Pure Yang-Mills Theory}

With the variables $E$ and $\tilde{E}$ introduced in the two preceding
sections, let us now examine the dual properties of the pure Yang-Mills
theory.  Since the theory in the standard (direct) formulation has a local
potential $A_\mu(x)$, it follows that if the theory is symmetric under
the dual transform (\ref{newduality}) introduced above, then there must
also be a local potential $\tilde{A}_\mu(x)$ in the dual formulation.  Now,
in the abelian theory, it was the equation of motion (\ref{gausslaw0})
which guaranteed via the Poincar\'e lemma the existence of the dual potential
$\tilde{A}_\mu(x)$; so we can hope that here too in the nonabelian theory,
it is the Yang-Mills equation of motion, namely (\ref{Gausslaw0}), which
guarantees the existence of the local potential $\tilde{A}_\mu(x)$.
We shall now show that this is indeed the case.

According to Polyakov \cite{Polyakov}, the Yang-Mills equation
(\ref{Gausslaw0}) can be written in terms of the loop variables
$F_\mu[\xi|s]$ as:
\begin{equation}
\delta^\mu(s) F_\mu[\xi|s] = 0.
\label{Polyakov}
\end{equation}
By (\ref{divEinF}) it follows that
\begin{equation}
\delta^\mu(s) E_\mu[\xi|s] = 0.
\label{PolyakovE}
\end{equation}
Hence by (\ref{curlEtilde}) the dual variables $\tilde{E}_\mu[\eta|t]$ have
to satisfy the condition:
\begin{equation}
\delta_\nu(t) \tilde{E}_\mu[\eta|t] - \delta_\mu(t) \tilde{E}_\nu[\eta|t] = 0.
\label{GausslawEd}
\end{equation}
However, we know from Section 2 that this is exactly the
condition for these variables to possess a local potential.  Indeed,
according to the arguments there, (\ref{GausslawEd}) implies the existence
of a $\tilde{W}[\eta|t]$ such that:
\begin{equation}
\tilde{E}_\mu[\eta|t] = \delta_\mu(t) \tilde{W}[\eta|t],
\label{Wxisd}
\end{equation}
and the local potential $\tilde{A}_\mu(x)$ is given by the dual analogue of
(\ref{AmuxinW}):
\begin{equation}
\tilde{A}_\mu(\eta(t)) \dot{\eta}^\mu(t) = \lim_{\epsilon \rightarrow 0}
   \tilde{W}[\eta|t].
\label{AmuxinWd}
\end{equation}
One sees thus that the existence of a local dual potential $\tilde{A}_\mu(x)$
is indeed guaranteed.

{}From previous work \cite{Chanstsou,Chantsou,Chanftsou2,Chanftsou1}, we
have learned that it is possible, and in fact even convenient for deriving
the dynamics of colour charges, to reformulate the Yang-Mills theory in
terms of loop variables.  This was done for the Polyakov
variables $F_\mu[\xi|s]$.  Let us do it now in terms of the variables
$E_\mu[\xi|s]$.  We have shown already in Section 2 that they
give a complete description of the theory although they have to be
constrained by the curl-free condition (\ref{GausslawE}).  Suppose
then we start with the standard Yang-Mills action:\footnote{For ${\it su}(2)$,
our convention is: $B = B^it_i, t_i = \tau_i/2, \,\mbox{\rm Tr} B = 2 \times$
sum of diagonal elements, so that $\,\mbox{\rm Tr}(t_it_j) = \delta_{ij}$.
Our results are given explicitly for ${\it su}(2)$ although they can be
trivially extended to any ${\it su}(N)$.}
\begin{equation}
{\cal A}_F^0 = -\frac{1}{16\pi} \int d^4x \,\mbox{\rm Tr} \{F_{\mu\nu}(x)
   F^{\mu\nu}(x)\},
\label{calA0F}
\end{equation}
which in terms of the Polyakov variables $F_\mu[\xi|s]$ takes the familiar
form:
\begin{equation}
{\cal A}_F^0 = -\frac{1}{4\pi \bar{N}} \int \delta \xi ds
   \,\mbox{\rm Tr}\{F_\mu[\xi|s] F^\mu[\xi|s]\} \dot{\xi}^{-2}(s),
\label{calA0FL}
\end{equation}
we have from (\ref{Emuxis}) in terms of $E_\mu[\xi|s]$:
\begin{equation}
{\cal A}_F^0 = -\frac{1}{4\pi \bar{N}} \int \delta \xi ds
   \,\mbox{\rm Tr} \{E_\mu[\xi|s] E^\mu[\xi|s]\} \dot{\xi}^{-2}(s).
\label{calA0FE}
\end{equation}
Incorporating the constraint (\ref{GausslawE}) into the action by means
of Lagrange multipliers $W_{\mu\nu}[\xi|s]$, we obtain:
\begin{equation}
{\cal A}_F = {\cal A}_F^0 + \int \delta \xi ds \,\mbox{\rm Tr}
   \{W^{\mu\nu}[\xi|s]
   (\delta_\nu(s) E_\mu[\xi|s] - \delta_\mu(s) E_\nu[\xi|s])\},
\label{calAF}
\end{equation}
the extremization of which with respect to the variables $E_\mu[\xi|s]$
yields then the equation of motion in parametric form:
\begin{equation}
E_\mu[\xi|s] = - (4\pi \bar{N} \dot{\xi}^2(s)) \delta^\nu(s) W_{\mu\nu}[\xi|s].
\label{PolyakovEp}
\end{equation}
The parameter $W_{\mu\nu}[\xi|s]$ being antisymmetric in its indices
$\mu,\nu$, (\ref{PolyakovEp}) is easily seen to imply (\ref{Polyakov}), or
in other words the Yang-Mills equation (\ref{Gausslaw0}) as expected.

Now earlier work has shown that the Lagrange multipliers in such a formulation
often play the role of a dual potential \cite{Chanftsou1}.  If so, we
expect that the dual potential $\tilde{A}_\mu(x)$ should be expressible
in terms of the parameters $W_{\mu\nu}[\xi|s]$.  For reasons which will be
made clear later when we deal with colour charges, we anticipate that
$\tilde{A}_\mu(x)$ is expressible in terms of $W_{\mu\nu}[\xi|s]$ as:
\begin{equation}
\tilde{A}_\mu(x) = - 8\pi \int \delta \xi ds \epsilon_{\mu\nu\rho\sigma}
   \omega(\xi(s))W^{\rho\sigma}[\xi|s] \omega^{-1}(\xi(s)) \dot{\xi}^\nu(s)
   \dot{\xi}^{-2} \delta(\xi(s)-\eta(t)).
\label{AdinWmunu}
\end{equation}
However, we have already given a formula for $\tilde{A}_\mu(x)$ in terms
of $\tilde{W}[\eta|t]$ in (\ref{AmuxinWd}).  To see that these
two expressions agree, substitute the expression (\ref{PolyakovEp})
above into the dual transform (\ref{newduality}) obtaining:
\begin{equation}
\omega^{-1}(\eta(t)) \tilde{E}_\mu[\eta|t] \omega(\eta(t))
   = 8 \pi \epsilon_{\mu\nu\rho\sigma} \dot{\eta}^\nu(t)
   \int \delta \xi ds \delta_\alpha(s) W^{\rho\alpha}[\xi|s]
   \dot{\xi}^\sigma(s) \delta(\xi(s)-\eta(t)),
\label{EdinWmunu}
\end{equation}
where for:
\begin{equation}
\mbox{\mbox{}$^*\!$} W_{\mu\nu}[\xi|s] = -\mbox{\small $\frac{1}{2}$}
   \epsilon_{\mu\nu\rho\sigma} W^{\rho\sigma}[\xi|s],
\label{Wmunustar}
\end{equation}
one can rewrite:
\begin{equation}
\epsilon_{\mu\nu\rho\sigma} \delta_\alpha(s) W^{\rho\alpha}[\xi|s]
   = -\{\delta_\mu(s) \mbox{\mbox{}$^*\!$} W_{\nu\sigma}[\xi|s] + \delta_\nu(s)
   \mbox{\mbox{}$^*\!$} W_{\sigma\mu}[\xi|s] + \delta_\sigma(s)
   \mbox{\mbox{}$^*\!$} W_{\mu\nu}[\xi|s]\}.
\label{difWinWstar}
\end{equation}
However, since loop quantities by definition have only loop derivatives
transverse to the loop, the last two terms inside the bracket on the
right-hand side of (\ref{difWinWstar}) give zero contributions when
substituted into (\ref{EdinWmunu}) giving:
\begin{equation}
\omega^{-1}(\eta(t)) \tilde{E}_\mu[\eta|t] \omega(\eta(t)) =
   - 8\pi \delta_\mu(t) \int \delta \xi ds \dot{\eta}^\nu(t)
   \mbox{\mbox{}$^*\!$} W_{\nu\sigma}[\xi|s] \dot{\xi}^\sigma(s)
   \delta(\xi(s) - \eta(t)),
\label{EdinWstar}
\end{equation}
where we have performed an integration by parts with respect to $\delta \xi$.
It follows then from (\ref{Wxisd}) that, apart from a constant term:
\begin{equation}
\omega^{-1}(\eta(t)) \tilde{W}[\eta|t] \omega(\eta(t))
   = - 8\pi \dot{\eta}^\mu(t) \int \delta\xi ds \epsilon_{\mu\nu\rho\sigma}
   W^{\rho\sigma}[\xi|s] \dot{\xi}^\nu(s) \delta(\xi(s)-\eta(t)),
\label{WdinWmunu}
\end{equation}
from which we obtain easily through (\ref{AmuxinWd}) the relation
(\ref{AdinWmunu}) as desired.

The structure of the preceding arguments is set out on the left-hand side of
Chart I, where the $\tilde{U}$-invariance will be demonstrated later.
The similarity with Chart I of \cite{Chanftsou1} for the abelian case
is obvious.

Next, we explore whether a similar structure is also obtained if we go over
into the dual formulation in terms of $\tilde{E}$.  Substituting the
expression (\ref{invertdual}) for $E$ in terms of $\tilde{E}$ into the
action ${\cal A}^0_F$ in (\ref{calA0FE}), we obtain on integrating over
$\xi$ and summing over indices:
\begin{equation}
{\cal A}^0_F = \frac{1}{4\pi \bar{N}} \int \delta \eta dt \,\mbox{\rm Tr}
   \{\tilde{E}_\mu[\eta|t] \tilde{E}^\mu[\eta|t]\} \dot{\eta}^{-2}(t),
\label{calA0FEd}
\end{equation}
where we have used the fact that $\tilde{E}_\rho[\eta|t]$ has only components
transverse to the loop $\eta$.  Apart from a sign, this is formally the
same as the action (\ref{calA0FE}) in terms of $E$.  Hence, if we extremize
this action under the constraint (\ref{GausslawEd}) ensuring that $\tilde{E}$
is curl-free to remove the redundancy of these variables, we see that the
problem will formally be exactly the same as for the direct formulation in
terms of $E$, producing the structure shown on the right-hand side of Chart I.
In other words, one has an exact dual symmetry as hoped.

\setcounter{equation}{0}
\section{Yang-Mills Theory with Charges}

Monopoles in gauge theories have by virtue of their topological nature an
intrinsic interaction with the gauge field, and Wu and Yang \cite{Wuyang} have
suggested a criterion whereby equations of motion for monopoles can be derived
as consequences of the topology without introducing an explicit interaction
term into the action.  The criterion has already been repeatedly applied with
success in earlier work \cite{Chanstsou,Chanftsou2,Chanftsou1}.   In case a
theory is dual symmetric, then both electric and magnetic charges are
monopoles in the appropriate fields so that the Wu-Yang criterion can be
applied to both giving dual symmetric equations as the result.  This was
the case in the abelian theory, and since we now claim that the Yang-Mills
theory is symmetric under the new generalized duality, it should be true
here also, which is what we wish now to demonstrate.

Let us start with a colour magnetic charge which is a monopole in the
Yang-Mills field, appearing as a topological obstruction with nontrivial
loop space holonomy, or equivalently non-zero loop space curvature
$G_{\mu\nu}[\xi|s]$, constructed from the Polyakov variable $F_\mu[\xi|s]$ as
connection \cite{Chanstsou,Chantsou}. This in turn means non-zero curl for
$E_\mu[\xi|s]$.  The statement that there is a classical (colour) magnetic
point charge $\tilde{g}$ moving along a world-line $Y^\mu(\tau)$ can thus
be explicitly expressed as:
\begin{equation}
\delta_\nu(s) E_\mu[\xi|s] - \delta_\mu(s) E_\nu[\xi|s]
   = - 4\pi J_{\mu\nu}[\xi|s],
\label{GausslawEJ}
\end{equation}
with:
\begin{equation}
J_{\mu\nu}[\xi|s] = \tilde{g} \epsilon_{\mu\nu\rho\sigma} \int d\tau
   {\cal K}(\tau) \frac{dY^\rho(\tau)}{d\tau} \dot{\xi}^\sigma(s)
   \delta(\xi(s)-Y(\tau)),
\label{Jmunuc}
\end{equation}
where ${\cal K}(\tau)$ is an algebra-valued quantity satisfying the conditiom
$\exp i\pi{\cal K} = -1$. \cite{Chanstsou}

The Wu-Yang criterion stipulates that equations of motion are to be derived
by imposing this definition  (\ref{GausslawEJ}) of the monopole as a constraint
on the free action, which is for the classical point particle:
\begin{equation}
{\cal A}^0 = {\cal A}^0_F - m \int d\tau.
\label{calA0c}
\end{equation}
Incorporating then the constraint (\ref{GausslawEJ}) by means of Lagrange
multipliers $W_{\mu\nu}[\xi|s]$ into the action, we have:
\begin{equation}
{\cal A} = {\cal A}^0 + \int \delta \xi ds \,\mbox{\rm Tr} [W^{\mu\nu}[\xi|s]
   \{\delta_\nu(s) E_\mu[\xi|s] - \delta_\mu(s) E_\nu[\xi|s]
   + 4\pi J_{\mu\nu}[\xi|s]\}].
\label{calA}
\end{equation}
We notice that at every space-time point not on the
world-line $Y^\mu(\tau)$ of the monopole, the condition (\ref{GausslawEJ})
says that the curl of $E$ vanishes, which is exactly the constraint we need
to impose on the $E$ variables to remove their intrinsic redundancy.  Hence,
in the action (\ref{calA}), where this constraint has already been
incorporated, $E_\mu[\xi|s]$ can now be taken as independent variables.

Extremizing then ${\cal A}$ in (\ref{calA}) with respect to the variables
$E_\mu[\xi|s]$ and $Y^\mu(\tau)$, we obtain again (\ref{PolyakovEp}) together
with:
\begin{eqnarray}
m \frac{d^2Y^\mu(\tau)}{d\tau^2} & = & - 8\pi \tilde{g} \int \delta\xi ds
   \epsilon^{\mu\nu\rho\sigma} \delta^\lambda(s) \,\mbox{\rm Tr}
   \{W_{\lambda\rho}[\xi|s] {\cal K}(\tau)\} \nonumber \\
   & & \times \frac{dY_\nu(\tau)}{d\tau} \dot{\xi}_\sigma(s)
   \delta(\xi(s)-Y(\tau)).
\label{Wong1}
\end{eqnarray}
{}From these the Lagrange multipliers $W_{\mu\nu}[\xi|s]$ can be eliminated
giving the Polyakov equation (\ref{Polyakov}) or (\ref{PolyakovE}) together
with:
\begin{eqnarray}
m \frac{d^2Y^\mu(\tau)}{d\tau^2}  & = & \frac{2 \tilde{g}}{\bar{N}}
   \int \delta\xi ds \epsilon^{\mu\nu\rho\sigma} \,\mbox{\rm Tr}
   \{E_\rho[\xi|s] {\cal K}(\tau)\}   \nonumber \\
&  & \times \dot{\xi}_\sigma(s) \dot{\xi}^{-2}(s) \frac{dY_\nu(\tau)}{d\tau}
   \delta(\xi(s)-Y(\tau)),
\label{Wong2}
\end{eqnarray}
where one sees that $E_\rho[\xi|s]$ appears in the combination:
\begin{equation}
\frac{2}{\bar{N}} \int \delta\xi ds \epsilon^{\mu\nu\rho\sigma} E_\rho[\xi|s]
   \dot{\xi}_\sigma(s) \dot{\xi}^{-2}(s) \delta(\xi(s)-Y(\tau)),
\label{Wong3}
\end{equation}
which is exactly what appeared also in the dual transform (\ref{newduality})
if one takes there the zero segmental width limit ($\epsilon \rightarrow 0$)
and put $\eta(t) = Y(\tau)$.  However, the other field equation of motion
(\ref{PolyakovE}) has already been shown via the dual transform to imply
the existence of a local gauge potential $\tilde{A}_\mu(x)$ for
$\tilde{E}_\mu[\eta|t]$, so that by (\ref{FmunuxinE}) in the limit of zero
segmental width:
\begin{equation}
\tilde{E}_\mu[\eta|t] \longrightarrow \tilde{F}_{\mu\nu}(\eta(t))
   \dot{\eta}^\nu(t),
\label{Edinx}
\end{equation}
with:
\begin{equation}
\tilde{F}_{\mu\nu}(x) = \partial_\nu \tilde{A}_\mu(x) - \partial_\mu
   \tilde{A}_\nu(x) + i \tilde{g} [\tilde{A}_\mu(x),\tilde{A}_\nu(x)].
\label{Fmunud}
\end{equation}
Whence, it follows that (\ref{Wong2}) reduces to:
\begin{equation}
m \frac{d^2Y^\mu(\tau)}{d\tau^2} = - \tilde{g} \,\mbox{\rm Tr} \{K(\tau)
   \tilde{F}_{\mu\nu}(Y(\tau))\} \frac{dY_\nu(\tau)}{d\tau},
\label{Wongd}
\end{equation}
with:
\begin{equation}
K(\tau) = \omega(Y(\tau)) {\cal K}(\tau) \omega^{-1}(Y(\tau)),
\label{Ktau}
\end{equation}
and $\tilde{F}_{\mu\nu}(Y(\tau))$ as given by (\ref{Fmunud}), which is the
dual of the Wong equation\footnote{This equation (\ref{Wongd}) should be
clearly distinguished from the equation with $\mbox{\mbox{}$^*\!$}{F}_{\mu\nu}
(x)$ in place of the $\tilde{F}_{\mu\nu}(x)$ here which we used to write in
previous work \cite{Chanstsou,Chanftsou2,Chanftsou1} prefaced by a warning that
it was meant only as illustration and should not be taken literally because
$\mbox{\mbox{}$^*\!$}{F}_{\mu\nu}(x)$ is patched and cannot be given a meaning
at the position $Y(\tau)$ of the monopole.   The present equation (\ref{Wongd})
does not suffer from these faults since $\tilde{F}_{\mu\nu}(x)$ is covariant
with respect to $\tilde{U}$- but invariant with respect to $U$-transformations
so that in the presence of the magnetic charge (which is a monopole of $E$
but only a source of $\tilde{E}$ ) it need not be patched at all and can
exist even at the position $Y(\tau)$ of the magnetic charge, just as in
the dual situation the Yang-Mills field $F_{\mu\nu}(x)$ requires no patching
when only electric charges are present.  Whatever patching that was needed
has been absorbed into the transformation matrix $\omega(x)$ which has itself
to be patched in the presence of the magnetic charge, as was shown in Section 6
of \cite{Chanftsou1}.  One notes further that the appearance of
$\mbox{\mbox{}$^*\!$}{F}_{\mu\nu}(x)$ in (\ref{Wongd}) instead of
$\tilde{F}_{\mu\nu}(x)$ would make the equation non-dual-symmetric since
according to Gu and Yang \cite{Guyang} a ``dual potential'' to
$\mbox{\mbox{}$^*\!$}{F}_{\mu\nu}(x)$ sometimes cannot exist.  On the other
hand, by virtue of the Yang-Mills equation or (\ref{PolyakovE}), a potential
for $\tilde{F}_{\mu\nu}(x)$ is known to exist through the arguments in
Section 2, thus restoring the symmetry with $F_{\mu\nu}(x)$ which is
endowed with a potential right from the beginning of the standard (direct)
formulation.  Technically, what had gone wrong in ``deriving'' the old
equation with $\mbox{\mbox{}$^*\!$}{F}_{\mu\nu}(Y(\tau))$ was that
one had to take first the limit of the segmental width $\epsilon \rightarrow 0$
and apply the formula (\ref{FmunuxinE}) in the expression (\ref{Wong3}) before
performing the integral, whereas the rule of the game as we understand it now
requires that the integral has to be first performed before the $\epsilon
\rightarrow 0$ limit is taken, a rule to which we have now adhered.}
\cite{Wong}.

Conversely, if we start with a colour electric charge considered as
a monopole of $\tilde{E}_\mu[\eta|t]$, we will obtain via exactly
the same arguments the dual of the above equations, namely:
\begin{equation}
\delta^\mu(t) \tilde{E}_\mu[\eta|t] = 0,
\label{PolyakovEd}
\end{equation}
which guarantees the existence of the potential $A_\mu(x)$ and is equivalent
to the ``dual Yang-Mills equation'':
\begin{equation}
\tilde{D}^\nu \tilde{F}_{\mu\nu}(x) = 0,
\label{Gausslaw0d}
\end{equation}
with:
\begin{equation}
\tilde{D}_\nu = \partial_\nu - i \tilde{g} [\tilde{A}_\nu(x), \ \ \ ],
\label{covdivd}
\end{equation}
together with the Wong equation:
\begin{equation}
m \frac{d^2Y^\mu(\tau)}{d\tau^2} = -g \,\mbox{\rm Tr} \{I(\tau)
   F^{\mu\nu}(Y(\tau))\} \frac{dY_\nu(\tau)}{d\tau}.
\label{Wong}
\end{equation}
The dynamics of a classical point charge is thus seen to be entirely dual
symmetric.

Consider next a Dirac particle carrying a colour magnetic charge.  The logical
steps for deriving its equations of motion in the gauge field using the
Wu-Yang criterion are the same as for the classical point particle, except
that the free action ${\cal A}^0$ is now: \cite{Chanftsou2,Chanftsou1}
\begin{equation}
{\cal A}^0 = {\cal A}^0_F + \int d^4x \bar{\psi}(x) (i \partial_\mu \gamma^\mu
   - m) \psi(x),
\label{calA0q}
\end{equation}
and the ``current'' $J_{\mu\nu}[\xi|s]$ in (\ref{calA}) is now the quantum
current:
\begin{equation}
J_{\mu\nu}[\xi|s] = \tilde{g} \epsilon_{\mu\nu\rho\sigma} \{\bar{\psi}(\xi(s))
   \omega(\xi(s)) \gamma^\rho t^i \dot{\xi}^\sigma(s) \omega^{-1}(\xi(s))
   \psi(\xi(s))\} t_i,
\label{Jmunuq}
\end{equation}
both depending on the wave function $\psi(x)$ of the particle.  Extremizing
the action (\ref{calA}) with respect to $E_\mu[\xi|s]$ yields again the
equation (\ref{PolyakovEp}) which is equivalent to the Polyakov equation
(\ref{Polyakov}) or the Yang-Mills equation (\ref{Gausslaw0}).  Extremizing
${\cal A}$ with respect to $\bar{\psi}(x)$ on the other hand yields:
\begin{equation}
(i \partial_\mu \gamma^\mu - m) \psi(x) = - \tilde{g} \tilde{A}_\mu(x)
   \gamma^\mu \psi(x),
\label{Diracd}
\end{equation}
where $\tilde{A}_\mu(x)$ is as given in (\ref{AdinWmunu}) and has already been
shown there to be the same as the dual potential.  This equation is thus
exactly
the dual of the Yang-Mills-Dirac equation for $\psi(x)$.

Starting with a colour electric charge considered as a monopole of
$\tilde{E}[\eta|t]$ and following exactly the same arguments will lead
easily to the dual equations to the above, namely the condition
(\ref{GausslawEd}) which guarantees the existence of the local gauge potential
$A_\mu(x)$ together with the Yang-Mills-Dirac equation for $\psi(x)$:
\begin{equation}
(i \partial_\mu \gamma^\mu - m) \psi(x) = - g A_\mu(x) \gamma^\mu \psi(x).
\label{Dirac}
\end{equation}
We have thus also for the quantum particle exact dual symmetry as we had hoped.

The result in this section is summarized in Chart II, which is seen to be
quite symmetric on left and right and entirely analogous to the Chart II
of \cite{Chanftsou1} for electrodynamics.

\setcounter{equation}{0}
\section{$U \times \tilde{U}$ Invariance}

That there is a dual doubling of the gauge symmetry in Yang-Mills theory
has already been shown previously \cite{Chanftsou2,Chanftsou1}.  Our task
here is merely to outline how this gauge symmetry operates in terms of
the new formulation, which turns out in fact to be considerably simpler
than it has appeared before.

Under simultaneous infinitesimal $U$ and $\tilde{U}$ local transformations
parametrized respectively by the gauge parameters $\Lambda(x)$ and
$\tilde{\Lambda}(x)$, the variables $E_\mu[\xi|s]$ and $\tilde{E}_\mu[\eta|t]$
transform as:
\begin{equation}
E_\mu[\xi|s] \longrightarrow [1 + ig \Lambda(\xi(s))] E_\mu[\xi|s]
   [1 - ig \Lambda(\xi(s))],
\label{Etrans}
\end{equation}
\begin{equation}
\tilde{E}_\mu[\eta|t] \longrightarrow [1 + i\tilde{g} \tilde{\Lambda}(\eta(t))]
   \tilde{E}_\mu[\eta|t] [1 - i \tilde{g} \tilde{\Lambda}(\eta(t))],
\label{Edtrans}
\end{equation}
while the rotation matrix $\omega(x)$ transforms as:
\begin{equation}
\omega(x) \longrightarrow [1 + i\tilde{g} \tilde{\Lambda}(x)] \omega(x)
   [1 - ig \Lambda(x)].
\label{omegatrans}
\end{equation}
It is clear then that the dual transform (\ref{newduality}) and its
inverse (\ref{invertdual}) are both gauge covariant.  Further, recalling that
the gauge parameters $\Lambda(\xi(s))$ and $\tilde{\Lambda}(\xi(s))$, being
local quantities, have zero loop derivatives (see the footnote in Section 3),
one sees that the relation (\ref{curlEtilde}) giving the curl of $\tilde{E}$
in terms of the divergence of $E$ which is so crucial for our duality
arguments is also gauge covariant.  That being the case, we need henceforth
consider the invariant properties for only one half of the dual symmetric
Charts I and II, since those for the other half will follow automatically.

Consider first Chart I for pure Yang-Mills fields. It is obvious that the
free field term in the action (\ref{calAF}) is gauge invariant.  The only
question then is how the Lagrange multipliers $W_{\mu\nu}[\xi|s]$ in the
constraint term will transform.  We put:
\begin{equation}
W_{\mu\nu}[\xi|s] \longrightarrow [1 + ig \Lambda(\xi(s))] \{W_{\mu\nu}[\xi|s]
   + i \tilde{g} \epsilon_{\mu\nu\rho\sigma} \delta^\rho(s)
   \tilde{\Lambda}^\sigma[\xi|s]\} [1 - ig \Lambda(\xi(s))],
\label{Wtrans}
\end{equation}
where we notice that in addition to a $U$-gauge rotation there is an
inhomogenious $\tilde{U}$-term parametrized by a vector quantity
$\tilde{\Lambda}^\sigma[\xi|s]$.  Under a pure
$\tilde{U}$-transformation (i.e. for $\Lambda =0$ in (\ref{Wtrans}))
the transformation of $W_{\mu\nu}[\xi|s]$ is that of the tensor potential
\footnote{Indeed, the Yang-Mills action when formulated in loop space
(\ref{calA0FL}) is entirely analogous to the Freedman-Townsend action
with $W_{\mu\nu}[\xi|s]$ here playing the role of the Freedman-Townsend
tensor potential \cite{Freedsen,Chanftsou3}.} discovered some years ago
first in supersymmetry theory \cite{Hayashi}.  On substituting (\ref{Wtrans})
into the action (\ref{calAF}), the $U$-gauge rotation factors cancel,
while the extra increment due to $\tilde{\Lambda}^\sigma[\xi|s]$,
after an integration by parts with respect to $\xi$, is seen to vanish by
virtue of the identity satisfied by the curl of $E$, namely:
\begin{equation}
\epsilon^{\mu\nu\rho\sigma} \delta_\rho(s)(\delta_\nu(s) E_\mu[\xi|s]
   - \delta_\mu(s) E_\nu[\xi|s]) = 0,
\label{BianchiE}
\end{equation}
leaving thus the whole action invariant.

The Lagrange multiplier $W_{\mu\nu}[\xi|s]$, however, is related to the dual
potential $\tilde{A}_\mu(x)$ by the relation (\ref{AdinWmunu}) so that its
transformation in (\ref{Wtrans}) will induce a transformation in the dual
potential.  The result is:
\begin{equation}
\tilde{A}_\mu(x) \longrightarrow [1 + i\tilde{g} \tilde{\Lambda}(x)]
   \tilde{A}_\mu(x) [1 - i\tilde{g} \tilde{\Lambda}(x)]
   - 2i\tilde{g} \partial_\mu \int \delta\xi ds \tilde{\Lambda}_\nu[\xi|s]
   \dot{\xi}^\nu(s) \delta(\xi(s)-x),
\label{Adtrans}
\end{equation}
where we have used the fact that $\tilde{\Lambda}_\nu[\xi|s]$ has only
transverse derivatives and performed an integration by parts with respect
to $\xi$.  Hence, we see that $\tilde{A}_\mu(x)$ transforms as a gauge
potential should, if we put:
\begin{equation}
\tilde{\Lambda}(x) = -8\pi \int \delta\xi ds \tilde{\Lambda}_\nu[\xi|s]
   \dot{\xi}^\nu(s) \delta(\xi(s)-x).
\label{Lambdadx}
\end{equation}

Given that it is this dual potential $\tilde{A}_\mu(x)$ which is coupled to
the wave function $\psi(x)$ of the magnetic charge, it is clear then that
the action (\ref{calA}) on Chart II is also invariant when the above
transformations are coupled with the usual transformations for the Wong
``charge'':
\begin{equation}
K(\tau) \longrightarrow [1 + i\tilde{g} \tilde{\Lambda}(x)] K(\tau)
   [1 - i\tilde{g} \tilde{\Lambda}(x)],
\label{Ktrans}
\end{equation}
and for the wave function:
\begin{equation}
\psi(x) \longrightarrow [1 + i\tilde{g} \tilde{\Lambda}(x)] \psi(x).
\label{psitrans}
\end{equation}
This last observation then completes our task.

\setcounter{equation}{0}
\section{Concluding Remarks}

Compared with our earlier work \cite{Chanftsou1} the present paper has gone
further in yielding an actual dual symmetry which had previously eluded us
and in giving simpler derivations of the old results.  The basis for this
improvement is the dual transform of (\ref{newduality}) which allows one
to switch at will from one formulation of the theory to its dual.  In terms
of this language, our previous treatment is only a half-way house where
only part of the dual transform has been carried out.  Thus, for example,
the so-called dual potential $T_{\mu\nu}[\xi|s]$ of \cite{Chanftsou1},
which is essentially our $W_{\mu\nu}[\xi|s]$ here, has in the present
treatment to undergo a further transform, namely (\ref{AdinWmunu}) which
is analogous to (\ref{newduality}), in order to give the genuine dual
potential $\tilde{A}_\mu(x)$.  It is the realization of this step which
eventually reveals the full dual symmetry.

Since the relationship between the two treatments can be worked out, given
the relation (\ref{Emuxis}) between the variables $E_\mu[\xi|s]$ used here
and the Polyakov variables $F_\mu[\xi|s]$ adopted in the earlier paper, no
detailed comparison need be given\footnote{We note that, for convenience,
we have used the same symbols in some cases to denote related but not
identical quantities in the two papers, but this we think should not lead
to any confusion.}.  There is one point, however, concerning the phase factor
$\Phi_\xi(s_+,0)$ occuring only in \cite{Chanftsou1} which puzzled
us at first and deserves perhaps a mention.  The factor $\Phi_\xi(s_+,0)$
appeared first in \cite{Chanftsou1} in the defining constraint for
the ``magnetic'' current:
\begin{equation}
G_{\mu\nu}[\xi|s] = - 4\pi J_{\mu\nu}[\xi|s],
\label{GausslawF}
\end{equation}
where for a classical point charge we had:
\begin{equation}
J_{\mu\nu}[\xi|s] = \tilde{g} \kappa[\xi|s] \epsilon_{\mu\nu\rho\sigma}
   \int d\tau \frac{dY^\rho(\tau)}{d\tau} \dot{\xi}^\sigma(s)
   \delta(\xi(s)-Y(\tau)),
\label{Jmunucold}
\end{equation}
with:
\begin{equation}
\kappa[\xi|s] = \Phi_\xi^{-1}(s_+,0) {\cal K}(\tau) \Phi_\xi(s_+,0),
\label{kappaxis}
\end{equation}
and ${\cal K}(\tau)$ a local quantity, while for a Dirac point charge we had:
\begin{equation}
J_{\mu\nu}[\xi|s] = \tilde{g} \epsilon_{\mu\nu\rho\sigma}
   [\bar{\psi}(\xi(s)) \omega(\xi(s)) \gamma^\rho t^i \omega^{-1}(\xi(s))
   \psi(\xi(s))] \Phi^{-1}_\xi(s_+,0) t_i \Phi_\xi(s_+,0).
\label{Jmunuqold}
\end{equation}
These expressions differ from (\ref{Jmunuc}) and (\ref{Jmunuq}) of this paper
by the factor $\Phi_\xi(s_+,0)$ and its inverse, where we note that the
argument is $s_+$ and not $s$ as elsewhere in this paper.\footnote{In
\cite{Chanftsou1,Chanftsou2}, we had actually written $\omega(s_+)$
instead of $\omega(s)$ as we do here to indicate that it was not affected
by loop differentiation, but this is in fact unnecessary in view of the
footnote of Section 3.}  That these factors should be there in
(\ref{Jmunucold}) and (\ref{Jmunuqold}) for consistency but not in
(\ref{Jmunuc}) and (\ref{Jmunuq}) can be seen as follows.  The
loop space curvature $G_{\mu\nu}[\xi|s]$ as exhibited in (\ref{Gmunu})
satisfies the Bianchi identity:
\begin{equation}
\epsilon^{\mu\nu\rho\sigma} {\cal D}_\rho(s) G_{\mu\nu}[\xi|s] = 0,
\label{BianchiF}
\end{equation}
where ${\cal D}_\mu(s)$ denotes the ``covariant loop derivative'':
\begin{equation}
{\cal D}_\mu(s) = \delta_\mu(s) - ig [F_\mu[\xi|s], \ \ \ ].
\label{covderivL}
\end{equation}
Hence, the current $J_{\mu\nu}[\xi|s]$ on the right-hand side of
(\ref{GausslawF}) must also satisfy this identity, which it does if it
contains the factors $\Phi_\xi(s_+,0)$ and $\Phi^{-1}_\xi(s_+,0)$ as shown
in (\ref{Jmunucold}) and (\ref{Jmunuqold}), but will not do so without
these factors.  On the other hand, although in the equation (\ref{GausslawEJ})
which is the equivalent to (\ref{GausslawF}) in terms of $E_\mu[\xi|s]$, the
current must also satisfy a similar identity (\ref{BianchiE}), this involves
only the ordinary loop derivative $\delta_\mu(s)$, and not the covariant loop
derivative ${\cal D}_\mu(s)$.  The expressions (\ref{Jmunuc}) and
(\ref{Jmunuq}) have thus no need for the phase factors $\Phi_\xi(s_+,0)$
and $\Phi_\xi^{-1}(s_+,0)$.  This difference between the ``currents'' in
the two treatments means that the corresponding Lagrange multipliers,
namely $L_{\mu\nu}[\xi|s]$ in the old and $W_{\mu\nu}[\xi|s]$ in the new,
are also related by a conjugation with respect to $\Phi_\xi(s_+,0)$, from
which it follows that the dual potential $\tilde{A}_\mu(x)$ defined in
\cite{Chanftsou1}, in spite of appearances, is in fact identical to that
defined here in (\ref{AdinWmunu}).

The above observation serves as a further example for the delicate handling
often required in loop space operations, which we consider as a weakness of
the whole loop space approach.  Although we believe we have considerably
improved our understanding in the present work, sufficiently in fact to
clarify one or two subtle points such as that in the Wong equation noted
in the footnote of Section 5 which we have not been able to make clear
before, we still feel strongly the lack of a general calculus for handling
complex loop space operations, the construction of which however is
unfortunately beyond our present capability.

Apart from this reservation, we find the result of the present paper
rather gratifying in that it seems to have answered the long-standing
question whether there is a dual symmetry for Yang-Mills theory and
gives even an explicit, though rather complicated, transformation between
dual variables, which is being sought for in other duality contexts.  For
us in particular, it seems to have answered also a question that we have
been asking on and off for some years concerning the dynamical properties
of nonabelian monopoles.  The answer to this turns out to be staggeringly
simple, namely that monopole dynamics is the same as that described by the
standard theory for Yang-Mills sources, only formulated in the dual fashion.
In consequence, one need not enquire, at least at the classical field level
so far studied, whether the charges one sees in nature are sources or
monopoles unless both types exist, for otherwise there will be no way to
distinguish them.  This is a rather unexpected result in view of the fact
that sources and monopoles are initially conceived as very different objects,
the former being essentially algebraic and the latter topological, and that
the dynamics is determined here via the Wu-Yang criterion by the topology
in an entirely different fashion from the manner that interactions for
sources are usually introduced.

\noindent {\bf Acknowledment}

One of us (TST) thanks the Wingate Foundation for partial support, while
another (JF) thanks the Particle Theory Group of the Rutherford Appleton
Laboratory for hospitaility during her summer visit there when part of
this work was done.

\end{document}